\documentclass{iopjournal}

\usepackage{amsmath}
\usepackage{graphicx}
\usepackage{amsmath,amssymb}
\usepackage{bm}
\usepackage{color}
\usepackage{tikz}
\usepackage[T1]{fontenc}
\usepackage[utf8]{inputenc}
\usepackage{booktabs}

\begin{document}

\articletype{Article type} 

\title{Bose-Fermi Mapping in Hubbard Models at Imaginary Chemical Potential and Phase-Induced Fermionization}

\author{Evangelos G. Filothodoros$^1$\orcid{https://orcid.org/0000-0002-5898-7288}}

\affil{$^1$Physics Department, Aristotle University of Thessaloniki, Thessaloniki, Greece}

\email{efilotho@physics.auth.gr}


\begin{abstract}
We find a mapping between the attractive Fermi-Hubbard model and the repulsive Bose-Hubbard model at finite temperature and at imaginary chemical potential $\mu =i\theta$. We show, by using a large $N$-expansion, that the partition functions of the two models are related by a simple shift $\theta \to \theta + \pi$. This condition maps the BCS--BEC crossover of attractive fermions to a Bose--Fermi crossover (fermion-like occupation) of repulsive bosons. Central feature of this correspondence plays the thermal kernel $g(\beta E,\phi),$
whose analytic continuation $g_B(\beta E,\phi) = g_F(\beta E,\phi+\pi)$ governs the bosonic and fermionic sectors. Interestingly, we are able to find that the special angles $\phi = 2\pi/3,4\pi/3$ for fermions correspond to $\phi = \pi/3,5\pi/3$ for bosons, marking the boundaries of a universal thermal window. We further argue that the present mechanism shows that fermionization can occur at finite interaction strength through a thermodynamic effect induced by the imaginary chemical potential. This suggests that it is a new way of fermionization (not a change in statistics but a fermion-like behaviour) unlike the Tonks--Girardeau limit, where fermionization arises from an infinite repulsive interaction and anyonic or Floquet-engineered systems where transmutation emerges from modified statistics or dynamics. Essentially, the phase $\phi$ is a statistical parameter; by twisting the thermal phase, it generates fermion-like behaviour without hard-core constraints or infinite repulsion but only by using thermodynamics.
We derive the gap equation and number equation for the bosonic model, highlighting the role of the imaginary chemical potential as a statistical regulator. Our results provide a unified framework for understanding crossovers in interacting lattice systems.
\end{abstract}

\section{Introduction}

The Hubbard model is one of the most interesting models in condensed matter physics, because it captures the complex competition between kinetic energy and local interactions. Usually the two specific kinds of the model, the attractive Fermi-Hubbard model and the repulsive Bose-Hubbard model, are treated as two distinct theories since they describe fundamentally different physics: the first helps us to describe pairing and superfluidity, while the other captures Mott insulation and Bose-Einstein condensation \cite{Zwerger, Garg, Diener, Bauer, Capogrosso}. However, despite their different natures, both models exhibit rich crossover phenomena driven by temperature, interaction strength, and external fields. In this work we are trying to answer a fundamental question: whether these two systems, operating under different statistics, can be connected through a common mathematical framework?

Since the extension of fermion-boson mapping \cite{Fujita, Filothodoros, Filothodoros2, Bonkhoff, Pupillo} from zero to finite temperature is always a recurrent theme in high energy and condensed matter physics, we demonstrate an exact duality between the two models, at finite temperature and in the presence of an imaginary chemical potential $\mu = i\theta$. Although this potential is not directly observable in lattice theory, it corresponds to a background $U(1)$ gauge field creating a holonomy along the thermal circle, analogous to Aharonov-Bohm phases \cite{Kapustin}. This temporal gauge field picture connects our analysis to a variety of phenomena where particles acquire phases from moving around closed loops and helps us to connect statistical mechanics to topological phase effects.

At the heart of this work, as a key insight, is the relation between fermionic and bosonic Matsubara frequencies, which leads to an exact duality of the corresponding partition functions. Consequently, this duality ensures that the thermal kernels of the two models are related by a simple shift and a main result  is the emergence of universal thermal windows at specific angles at $\phi = \pi/3$ and $5\pi/3$ for the bosonic model.  This observation allows us to ask whether these boundaries map exactly the corresponding boundaries from the attractive Fermi-Hubbard model, and indeed this happens in a remarkable and symmetric way.

While at the Tonks--Girardeau limit \cite{Girardeau, Paredes, Bera, Muraev}, where bosons fermionize due to an infinitely strong repulsive interaction, our results suggest that a similar fermion-like behaviour can emerge at interaction of finite strength when the system is at an imaginary chemical potential. Having this idea in mind, we see that fermionization is not driven by a hard-core constraint on the wavefunction \cite{Maiti}, but by a redistribution of spectral weight. In this picture, the imaginary chemical potential suppresses low-energy occupation and enhances higher-energy modes within a finite window $\phi \in [\pi/3,5\pi/3]$ ($\pi/3$ or $\phi_{\text{flux}}=1/6$ is special according to \cite{Lelas, Boesl, Iskin} since this value appear in lattice bosonic systems with real magnetic flux, where it controls strongly correlated and topological states and as a lattice phase geometry, relative to the boundaries), producing an effective exclusion-like behaviour at the level of thermal occupations. This reveals a regime where, despite the lack of a formal Bose-Fermi mapping at the operator level \cite{Carr}, the thermal occupations mimic fermionic behaviour through an interplay of interaction and entropic pressure. On the other hand, dimensionality enters only through the density of states $\rho(\epsilon)$, allowing the framework to be extended to various lattice geometries in the future, including 1D chains, 2D square lattices (with van Hove singularities), and 2D honeycomb lattices (with Dirac cones).

This paper is organized as follows. In Section 2, we present the repulsive Bose-Hubbard model with an imaginary chemical potential and derive its effective action using a large-$N$ expansion. Section 3 derives the gap equation, number equation, and the fermionization-like window of the bosonic model. Section 4 examines the phase structure through the thermal kernel and the extrema of number $N_b$ and section 5 reviews the main results for the attractive Fermi-Hubbard model at imaginary chemical potential and large-$N$. Section 6 establishes the duality between the attractive Fermi and repulsive Bose models, including the mapping of thermal kernels. In Section 7 we derive a universal statistical transmutation framework for the two models and we conclude in Section 8 with a summary and conclusions. In Appendix we have some useful calculations.

\section{Repulsive Bose-Hubbard Model and Large-\(N\) Formulation}

We consider the spinless Bose-Hubbard model on a lattice at inverse temperature
$\beta$, with a fixed total boson number $N_b$, and $n_i = b_i^\dagger b_i$, so the canonical partition function is \cite{ Filothodoros, Filothodoros2, Kapustin, Citro}
\begin{equation}
Z_{N_b} = \mathrm{Tr}\left[ \delta(\hat N - N_b)\, e^{-\beta H} \right],
\qquad
\hat N = \sum_i n_i.
\end{equation}
Introducing the Fourier representation of the projector,
\begin{equation}
\delta(\hat N - N_b)
= \frac{1}{2\pi} \int_0^{2\pi} d\theta \;
e^{i\theta(\hat N - N_b)},
\end{equation}
the canonical partition function can be written as
\begin{equation}
Z_{N_b}
= \frac{1}{2\pi} \int_0^{2\pi} d\theta\;
e^{-i\theta N_b}\, Z(\theta),
\end{equation}
where $Z(\theta)$ is a grand-canonical partition function with an
imaginary chemical potential $\mu = i\theta$, for $\theta$ to be dimensionless. As we will see in next Section, the analysis is performed in the $\theta$-ensemble, where the boson number is not fixed but determined dynamically via
\begin{equation}
N_b(\theta) = \frac{\partial \ln Z(\theta)}{\partial (i\theta)}.
\end{equation}

\subsection{Particle-channel formulation and large-$N$ effective action}

To describe the Bose--Fermi crossover, the repulsive Bose-Hubbard interaction must be
decoupled in the \emph{particle channel}, since the formation of composite fermions
cannot be captured by a simple mean-field decoupling. We introduce a large-$N$ generalization by extending the bosons to $N$ flavors $\alpha=1,\dots,N$, which provides a controlled saddle-point limit. The momentum version of Hamiltonian is \cite{Vitoriano, Giering}
\begin{equation}
H
=
\sum_{\mathbf k,\alpha}
\epsilon_{\mathbf k}\,
b_{\mathbf k\alpha}^\dagger
b_{\mathbf k\alpha}
+
\frac{U}{2N}
\sum_i
\sum_{\alpha,\beta}
b_{i\alpha}^\dagger
b_{i\alpha}^\dagger
b_{i\beta}
b_{i\beta},
\end{equation}
where $U>0$ is the repulsive interaction. The factor $1/N$ ensures that
the interaction energy scales extensively in the large-$N$ limit.
The Euclidean action with imaginary chemical potential then takes the form
\begin{equation}
S=\int_0^\beta d\tau\Bigg[
\sum_{\mathbf k,\alpha}
b_{\mathbf k\alpha}^\dagger
(\partial_\tau+\epsilon_{\mathbf k}-i\theta)
b_{\mathbf k\alpha}
+
\frac{U}{2N}
\sum_{i,\alpha,\beta}
b_{i\alpha}^\dagger
b_{i\alpha}^\dagger
b_{i\beta}
b_{i\beta}
\Bigg].
\end{equation}
The quartic interaction is decoupled via a Hubbard--Stratonovich
transformation in the particle channel by introducing a complex bosonic
field $\Phi_i(\tau)$ (which will become the fermionic order parameter):
\begin{equation}
\exp\!\left[
-\frac{U}{2N}
\int d\tau\,
b^\dagger b^\dagger b b
\right]
=
\int \mathcal D\Phi\,\mathcal D\Phi^*\;
\exp\!\left[
-\int d\tau
\left(
\frac{N|\Phi|^2}{2U}
+
\Phi^* b b
+
\Phi b^\dagger b^\dagger
\right)
\right].
\end{equation}
After this transformation, the bosonic action becomes quadratic in the original boson fields.
Introducing the Nambu spinor for bosons,
\begin{equation}
\Upsilon_{\mathbf{k}\alpha} = 
\begin{pmatrix}
b_{\mathbf{k}\alpha} \\
b_{-\mathbf{k}\alpha}^\dagger
\end{pmatrix},
\end{equation}
the inverse bosonic Green's function is
\begin{equation}
\mathcal G^{-1}(i\omega_n,\mathbf k)=
\begin{pmatrix}
(i\omega_n+i\theta)-\epsilon_{\mathbf k} & -\Phi \\
\Phi^* & -(i\omega_n+i\theta)-\epsilon_{\mathbf k}
\end{pmatrix},
\end{equation}
where $\omega_n = 2\pi n/\beta$ are the bosonic Matsubara frequencies.
Its determinant yields the quasiparticle spectrum:
\begin{equation}
E_{\mathbf k}
=
\sqrt{\epsilon_{\mathbf k}^2 + |\Phi|^2},
\end{equation}
which governs the Bose--Fermi crossover physics.
The physical excitation spectrum is obtained from the poles of the Green's function. Since the bosonic action is quadratic, the fields can be
integrated out and the effective action is
\begin{equation}
S_{\mathrm{eff}}[\Phi,\theta]
=
\int_0^\beta d\tau
\frac{N|\Phi|^2}{2U}
+
N\,\mathrm{Tr}\,\ln \mathcal G^{-1}[\Phi,\theta],
\end{equation}
where the trace is taken over lattice sites, imaginary time (or
Matsubara frequencies), Nambu indices, and internal degrees of freedom.
We note the $+$ sign in front of the trace, which arises from the bosonic
functional integral (determinant in denominator).

The static saddle-point configuration is $\Phi_i(\tau) = \Phi$ and we may say that it is interpreted as an auxiliary field encoding local two-particle correlations $\Phi \sim \langle b_i b_i \rangle$ and also it does not correspond to a conventional order parameter for repulsive bosons. In contrast to the attractive fermionic case, where the analogous field describes Cooper pairing, here $\Phi$ measures the tendency of bosons to occupy the same site. Repulsive interactions suppress such configurations, so that the fermionization regime is characterized by a reduction of $\Phi$. Essentially, we could argue that $\Phi$ serves as an indicator of the degree of bosonic coherence versus fermion-like avoidance, rather than a true symmetry-breaking order parameter.

The trace over Nambu space yields
\begin{equation}
\det \mathcal G^{-1}(i\omega_n,\mathbf k)
=
-(i\omega_n +i\theta)^2 + \left(
\epsilon_{\mathbf k}^2 + |\Phi|^2
\right),
\end{equation}
The functional trace can therefore be written as
\begin{equation}
\mathrm{Tr}\,\ln \mathcal G^{-1}
=
\sum_{\mathbf k}
\sum_{n}
\ln\!\left[
-(i\omega_n+i\theta)^2 + E_{\mathbf k}^2
\right].
\end{equation}
Performing the Matsubara frequency sum using contour-integration gives
\begin{equation}
\sum_{n}
\ln\!\left[-(i\omega_n +i\theta)^2 + E_{\mathbf k}^2\right]
= \beta E_{\mathbf k}
+ \ln\!\left(1 - e^{-\beta(E_{\mathbf k}+i\theta)}\right)+ \ln\!\left(1 - e^{-\beta(E_{\mathbf k}-i\theta)}\right) + \text{constant}.
\end{equation}
The constant (independent of $\Phi$ and $\theta$) can be absorbed into the
normalization of the partition function.
Substituting this result back into the effective action yields
\begin{equation}
S_{\mathrm{eff}}[\Phi,\theta]
= N
\Big[
\frac{|\Phi|^2}{2U}
+\sum_{\mathbf k}
\Big(E_{\mathbf k}
+
\left[
\ln\!\left(1 - e^{-\beta(E_{\mathbf k}+i\theta)}\right)
+
\ln\!\left(1 - e^{-\beta(E_{\mathbf k}-i\theta)}\right)
\right]\Big)\Big].
\end{equation}
In the thermodynamic limit, the lattice momentum sum is replaced by an
integral over the first Brillouin zone,
\begin{equation}
\sum_{\mathbf k}
\;\longrightarrow\;
\int_{\mathrm{BZ}}
\frac{d^{d-1} k}{(2\pi)^{d-1}}.
\end{equation}
The effective action density then becomes
\begin{equation}
S_{\mathrm{eff}}
=
N
\Big[
\frac{|\Phi|^2}{2U}
+\int_{\mathrm{BZ}}
\frac{d^{d-1} k}{(2\pi)^{d-1}}
\Big(E_{\mathbf k}
+
\left[
\ln\!\left(1 - e^{-\beta(E_{\mathbf k}+i\theta)}\right)
+
\ln\!\left(1 - e^{-\beta(E_{\mathbf k}-i\theta)}\right)
\right]\Big)\Big].
\end{equation}

\section{Gap Equations for the Repulsive Bose-Hubbard Model}

The first gap equation of the model for the order parameter $\Phi$ is

\begin{equation}
\frac{\partial S_{\mathrm{eff}}}{\partial \Phi^*} = 0
\end{equation}
and yields the equation
\begin{equation}
\frac{1}{U}
=
\int_{\mathrm{BZ}}
\frac{d^{d-1} k}{(2\pi)^{d-1}}
\,
\frac{1}{E_{\mathbf k}}
\left[
1
+
f_B(E_{\mathbf k}+i\theta)
+
f_B(E_{\mathbf k}-i\theta)
\right],
\end{equation}
where 
$f_B(x) = \frac{1}{e^{\beta x} - 1}$ is the Bose-Einstein distribution function.
Note the plus signs and the Bose-Einstein statistics, which differ from the fermionic case. At zero temperature, $f_B(x) \to 0$, so this reduces to
\begin{equation}
\frac{1}{U}
=
\int_{\mathrm{BZ}}
\frac{d^{d-1} k}{(2\pi)^{d-1}}
\,
\frac{1}{E_{\mathbf k}}.
\end{equation}
For $\theta = 0$, we obtain the standard gap equation for the repulsive Bose-Hubbard model
\begin{equation}
\frac{1}{U}
=
\int_{\mathrm{BZ}}
\frac{d^{d-1} k}{(2\pi)^{d-1}}
\,
\frac{1 + 2f_B(E_{\mathbf k})}{E_{\mathbf k}}.
\end{equation}
The canonical constraint is enforced by the saddle-point condition
with respect to $\theta$:
\begin{equation}
\frac{\partial S_{\mathrm{eff}}}{\partial \theta} = 0.
\end{equation}
This yields the number equation \footnote{Note that the expression in (\ref{N_b}) is purely imaginary. The physical boson number is obtained by multiplying by $-i$, yielding a real quantity that measures the imbalance between adding and removing "fermionized" bosons.}
\begin{equation}
N_b
=
\int_{\mathrm{BZ}} \frac{d^{d-1} k}{(2\pi)^{d-1}} \,
\Big[
f_B(E_{\mathbf k}-i\theta) - f_B(E_{\mathbf k}+i\theta)
\Big],
\label{N_b}
\end{equation}
where $N_b$ is the total boson number. In the bosonic case it quantifies particle–hole asymmetry induced by the imaginary chemical potential and reflects fermionization-like regime, or in other words how much the system prefers particle-like versus hole-like excitations, analogous to the fermionic case. For $D=d-1=1$ we will find that the fermionization-like window is where this asymmetry is maximal, corresponding to the point where bosons most closely mimic fermionic behaviour, since fermions have a natural particle-hole asymmetry due to the Pauli principle.

\paragraph{Fermionization-like window and spectral redistribution.}
Studying the phase structure of our theory in next section, we will find that at the special angles $\phi = \pi/3$ and $\phi = 5\pi/3$, the repulsive Bose--Hubbard model at imaginary chemical potential enters a regime of maximal spectral redistribution. Here the competition between repulsive interactions and phase frustration suppresses Bose condensation and enhances entropy. This results in an effective fermion-like occupation profile, characterized by a flattening of the energy distribution and a reduced tendency for bosons to macroscopically occupy low-energy states. Although there is no true Pauli principle in our case, the thermal kernel induces an emergent exclusion-like behaviour that mimics fermionic statistics at the level of occupation numbers. The two angles correspond to opposite occupations: $\phi = \pi/3$ enhances particle-like excitations, whereas $\phi = 5\pi/3$ produces a sign-reversed, hole-like response, reflecting a dual structure controlled by the imaginary chemical potential. This situation is analogous to Tonks-Girardeau (TG) gas, where bosons "fermionize" because they cannot pass through each other when they reach the limit of infinite repulsion energy $U\rightarrow \infty$. So we may say that our model suggests a different route to the same destination: while TG bosons require infinite energy to fermionize, our system achieves this state at finite interaction, driven by the 'statistical pressure' of the imaginary chemical potential. On the other hand, we may say that the point $\phi=\pi$ is not a maximally bosonic or fermionic configuration,
but rather a (particle-like)--(hole-like) symmetric point where the number
equation vanishes identically.

The gap equation may be split into two parts,
\begin{equation}
I_1 = \int_{\mathrm{BZ}}
\frac{d^{d-1} k}{(2\pi)^{d-1}}
\,
\frac{1}{E_{\mathbf k}}
\end{equation}
and 
\begin{equation}
I_2 = \int_{\mathrm{BZ}}
\frac{d^{d-1} k}{(2\pi)^{d-1}}
\,
\frac{1}{E_{\mathbf k}}
\left[
f_B(E_{\mathbf k}+i\theta)
+
f_B(E_{\mathbf k}-i\theta)
\right].
\end{equation}
Note the plus sign in front of $I_2$ (compared to the minus sign in the fermionic case).
For 1D chains at finite temperature, we take simply
\begin{equation}
I_1 = \int_{\mathrm{BZ}}
\frac{dk}{2\pi}\frac{1}{E_{\mathbf k}}
\end{equation}
and
\begin{equation}
I_2 = \int_{\mathrm{BZ}}
\frac{dk}{2\pi}\frac{1}{E_{\mathbf k}}\left[
f_B(E_{\mathbf k}+i\theta)
+
f_B(E_{\mathbf k}-i\theta)
\right].
\end{equation}
If we define the vacuum (critical) coupling by
\begin{equation}
\frac{1}{U_c} \equiv I_1(\Phi=0)
= \int_{\mathrm{BZ}}\frac{dk}{2\pi}\frac{1}{|\epsilon_k|},
\end{equation}
and subtract this value, the renormalized gap equation becomes
\begin{equation}
\frac{1}{U} - \frac{1}{U_c}
=
\int_{\mathrm{BZ}}\frac{dk}{2\pi}
\left(
\frac{1}{\sqrt{\epsilon_k^2 + |\Phi|^2}} - \frac{1}{|\epsilon_k|}
\right)
+ I_2(\Phi,\theta).
\end{equation}
The sign of $\frac{1}{U} - \frac{1}{U_c}$ controls the crossover:
positive corresponds to the Bose regime (weak repulsion), negative to the Fermi regime (strong repulsion, fermionization), and
$\frac{1}{U} = \frac{1}{U_c}$ marks the crossover (fermionization) point.

\section{Phase Structure: Bose, Fermionization, and Crossover Regimes}

\subsection{General Setup for Bosons}

Using the identity for Bose-Einstein distributions \cite{Roberge},
\begin{equation}
f_B(E_k+i\theta)+f_B(E_k-i\theta) = -1 - \frac{\sinh(\beta E_k)}{\cosh(\beta E_k) - \cos(\beta\theta)},
\end{equation}
and $\phi \equiv \beta\theta$ as a dimensionless variable, the bosonic gap equation becomes:
\begin{equation}
\delta_u \equiv \frac{1}{U} - \frac{1}{U_c}
= \int_{\mathrm{BZ}} \frac{dk}{2\pi} \left[ 
\frac{\sinh(\beta E_k)}{E_k\big[\cosh(\beta E_k)-\cos\phi\big]}
- \frac{1}{|\epsilon_k|}
\right].
\end{equation}
Note the minus sign in the denominator ($\cosh(\beta E_k) - \cos\phi$) compared to the fermionic case.
For a lattice dispersion $\epsilon_k = -2t\cos k$, the band minimum occurs at $k=0$ (or $k=\pi$ depending on the sign). Near the band minimum $k=0$, we expand:
\begin{equation}
\epsilon_k \approx -2t\left(1 - \frac{k^2}{2}\right) = -2t + t k^2,
\end{equation}
so the low-energy dispersion is quadratic:
\begin{equation}
\epsilon_k \approx \text{constant} + \frac{k^2}{2m^*},
\end{equation}
where $m^* = 1/(2t)$ is the effective mass. For simplicity, we measure energies from the band bottom, so $\epsilon_k \approx t k^2$.
If we split the integral into low-energy contributions near $k=0$ (within a cutoff $|k|<\Lambda$) and the remainder $C_{\mathrm{UV}}$ (higher momentum part inside BZ), we obtain the low-energy integral:
\begin{equation}
\delta_u = \underbrace{\int_{0}^{\Lambda} \frac{dk}{\pi} \left[ 
\frac{\sinh(\beta E_k)}{E_k\big[\cosh(\beta E_k)-\cos\phi\big]}
- \frac{1}{t k^2}
\right]}_{\displaystyle \mathcal{F}_{\mathrm{low}}(\Phi,T,\phi;\Lambda)}
\;+\; C_{\mathrm{UV}}(\Lambda).
\end{equation}
The critical coupling is defined as:
\begin{equation}
\frac{1}{U_c} = \int_{0}^{\Lambda} \frac{dk}{\pi} \frac{1}{t k^2}.
\end{equation}
Since $C_{\mathrm{UV}}(\Lambda)$ comes from the rest of the Brillouin zone, it is \textbf{independent of $\Phi$ and $T$} in the low‑energy limit ($\Lambda \ll 1$, low $T$ and $\Phi$). Essentially $C_{\mathrm{UV}}$ encodes the contribution of high energy modes within the lattice band where the phase dependence and the structure of thermal windows are governed almost entirely by the low energy part $\mathcal{F}_{\mathrm{low}}$. So we may absorb it into a redefinition of the parameter $\delta_u$.

\subsection{Bosonic Thermal Kernel}

The bosonic thermal kernel is defined as:
\begin{equation}
g_B(x, \phi) \equiv \frac{\sinh x}{\cosh x - \cos\phi}, \qquad x = \beta E_k.
\label{bosonic_kernel}
\end{equation}
Using the key identity
\begin{equation}
g_B(x,\phi) - 1 = \frac{e^{-x} - \cos\phi}{\cosh x - \cos\phi},
\end{equation}
we obtain:
\begin{equation}
g_B(x,\phi) \gtreqless 1 \quad \Longleftrightarrow \quad e^{-x} \gtreqless \cos\phi.
\end{equation}

Since $e^{-x} \le 1$ for $x \ge 0$, this inequality has interesting structure:
\begin{itemize}
\item If $\cos\phi = 1$ ($\phi = 0, 2\pi$), then $g_B(x,\phi) > 1$ for all $x > 0$
\item If $\cos\phi < 1$, there exists a threshold $x_* = -\ln(\cos\phi)$ where $g_B = 1$
\end{itemize}

At the special angle $\phi = \pi/3$ (and similarly at $5\pi/3$), where $\cos\phi = 1/2$, the kernel becomes:
\begin{equation}
g_B(x,\pi/3) = \frac{\sinh x}{\cosh x - \tfrac{1}{2}},
\end{equation}
and satisfies
\begin{equation}
g_B(x,\pi/3) - 1 = \frac{e^{-x} - \tfrac{1}{2}}{\cosh x - \tfrac{1}{2}}.
\end{equation}
This expression changes sign at $x_* = \ln 2$, exactly as in the fermionic case.
Consequently, the integrand of the bosonic gap equation exhibits a nontrivial structure: low-energy modes ($x < \ln 2$) contribute with $g_B < 1$, while high-energy modes ($x > \ln 2$) contribute with $g_B > 1$.
At the critical point $\delta_u = 0$, these contributions balance exactly after inclusion of the high-energy term $C_{\mathrm{UV}}$.
Therefore, the angles $\phi = \pi/3$ and $5\pi/3$ mark the points where the bosonic thermal kernel produces a sign-changing contribution across energy scales. This leads to a particularly sensitive balance in the gap equation at the critical coupling, corresponding to the \textbf{fermionization-like window} where bosons are under an effective exclusion principle at the level of thermodynamics, in order to minimize 
free energy via entropy gain and interaction avoidance:
\begin{equation}
\boxed{ \phi \in \left[ \frac{\pi}{3}, \frac{5\pi}{3} \right] },
\end{equation}
Notably, $ln2$ appears as entropy values of this order in \cite{Osolin} as a similar result. In the low-energy limit, the thermal kernel can be approximated as
\begin{equation}
g(\epsilon,\phi) \approx \frac{1}{(1-\cos\phi) + \frac{(\beta\epsilon)^2}{2}},
\end{equation}
so that at small $\epsilon$,
\begin{equation}
g(\epsilon,\phi)-1 \approx \frac{\cos\phi}{1-\cos\phi}.
\end{equation}
This shows that the sign of the low-energy contribution is determined 
by $\cos\phi$. 
Since the kernel is peaked at low energies, the integral 
$\int d\epsilon\, \rho(\epsilon)[g(\epsilon,\phi)-1]$ is dominated by 
this region, implying that the sign of the integral is approximately 
controlled by the sign of $g(\epsilon,\phi)-1$. Deviations from this simple 
picture arise from finite-energy contributions, which shift the precise 
location of the phase boundaries away from the condition $\cos\phi=0$.

\paragraph{Bosonic regimes.} 

For the repulsive Bose--Hubbard model at imaginary chemical potential,
the regimes are controlled by the parameter $\delta_u = 1/U - 1/U_c$.

For $\delta_u < 0$ (strong repulsion), we see that clearly bosons exhibit a fermionization
tendency: the system actively suppresses low-energy states, pushing spectral weight toward higher energies. This is not a true transition to fermionic statistics, but rather
an interaction-driven flattening of the occupation profile.

On the other hand for $\delta_u > 0$ (weak repulsion), the system remains in a Bose-like
regime, where occupation of low-energy states is enhanced and the phase
twist induces only mild distortions of the distribution.

At the critical point $\delta_u = 0$, the system undergoes a crossover
between these different regimes. The effect is strongest at the special angles
$\phi = \pi/3, 5\pi/3$, where the thermal kernel defines the fermionization window.

However, it is important to say that we should be careful to distinguish this large-$N$ formulation,
based on a particle-channel decoupling, from the classic Mott-insulator-to-superfluid transition, which is associated with the condensation of the single-particle
field $\langle b \rangle$.
Instead, the present approach focuses on a distinct crossover driven by
repulsive interactions and imaginary chemical potential, leading to an
effective fermionization of bosonic excitations.

In this sense, the regimes identified here should be interpreted as
``Bose-like'' and ``fermionized'' regimes, rather than true superfluid
or Mott insulating phases. The strong-repulsion regime ($\delta_u<0$)
shares qualitative features with Mott physics, such as suppression of
low-energy occupation, but does not exhibit a true gap associated with
integer filling.

At the particle--hole-like symmetric point $\phi = \pi$, the kernel reduces to
\[
\frac{1}{\cosh(\beta E)+1},
\]
which is identical to the fermionic thermal kernel. This establishes a precise thermodynamic correspondence, although the microscopic degrees of freedom remain bosonic.
Unlike the BCS--BEC crossover, where composite fermions create bound pairs that behave as bosons, our mechanism does not rely on pairing. Instead, the imaginary chemical potential modifies the thermal kernel, leading to fermion-like occupation constraints for elementary bosons without changing their underlying statistics.

On the other hand, as we have already mentioned, if we interpret $C_{\mathrm{UV}}$ as a renormalization shift of the interaction parameter, we may write
\begin{equation}
\delta_u^{\text{eff}}=\delta_u - C_{\mathrm{UV}} = \mathcal{F}_{\mathrm{low}}(\Phi,T,\phi),
\end{equation}
where $\mathcal{F}_{\mathrm{low}}$ encodes the low-energy contribution weighted by the thermal kernel $g(\epsilon,\phi)$.

\begin{figure}[h]
               \begin{center}
                \includegraphics[scale=0.44]{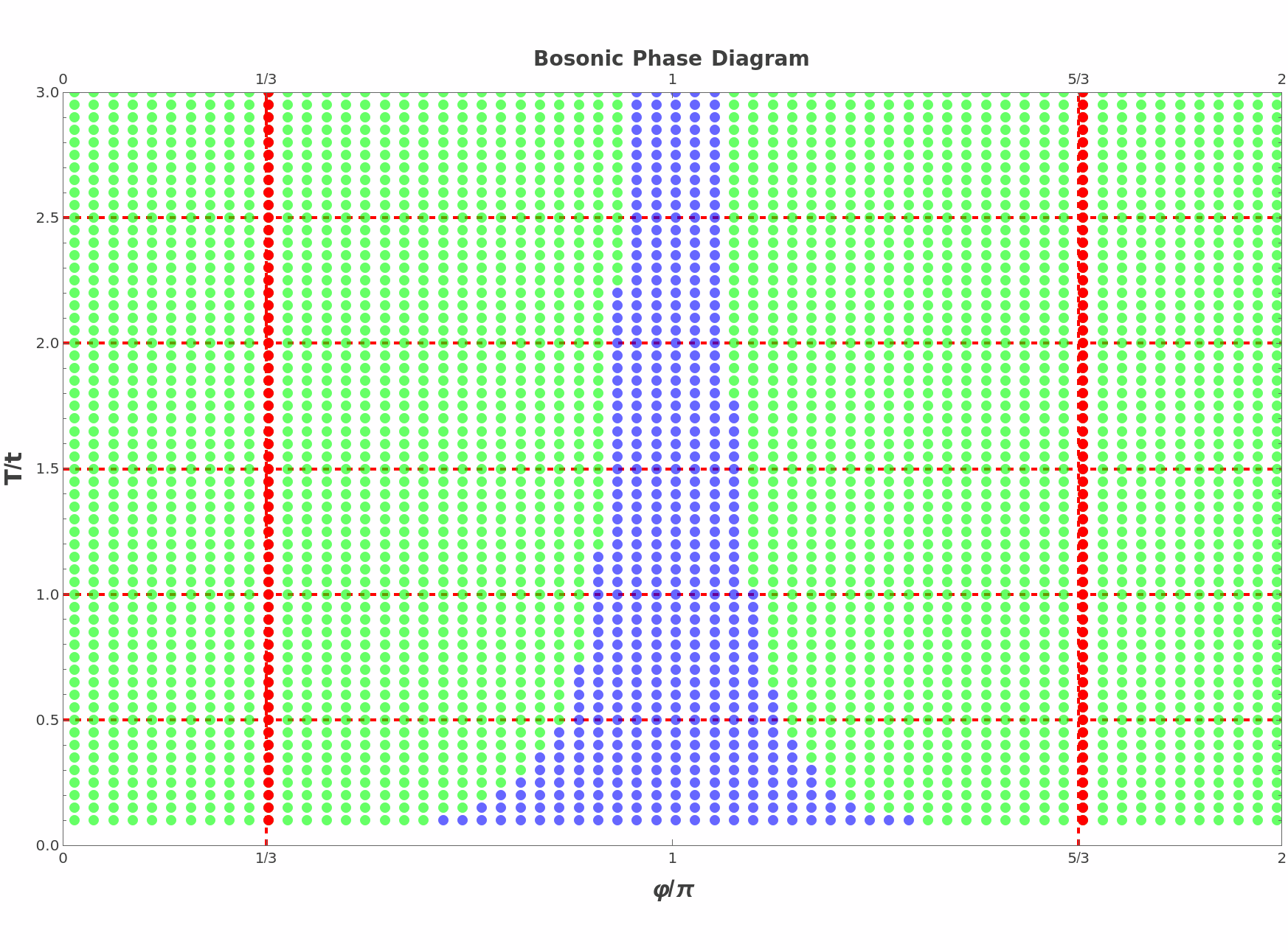}
                \end{center}
                 \caption{Bose-Fermi crossover for $F+C_{\text{UV}}$ for the Bose Hubbard model at imaginary chemical potential and $\delta_u=0$. Lower $T/t$ means entropy term $-TS$ becomes less negative so $U$ term dominates. Then the system spreads out to avoid repulsion and we have more fermion-like occupation. The blue region indicates enhanced fermionization ($g_B<1$), green indicates suppressed fermionization ($g_B>1$), and red dots at $\beta\theta=\pi/3,5\pi/3$ mark where $\Phi=0$ and $N_b$ is maximized. The parameters we have used in Mathematica plot are $t=1$, $\Lambda=0.5$, $C_{\text{UV}}=+5.2$. The magnitude and sign of the "ultraviolet" constant $C_{\mathrm{UV}}$ are determined by the
high-energy behaviour of the dispersion relation. For fermions, the low-energy dispersion
is linear, $\epsilon_k \sim k$, leading to a logarithmic divergence in the integral
$\int dk/k$. For bosons, the
dispersion is quadratic, $\epsilon_k \sim k^2$, which produces a much stronger linear
divergence $\int dk/k^2$. As a result, the raw integral inside the fermionization
window is strongly negative. To overcome this large negative contribution and reveal a well-balanced and visible blue ($g_B<1$) Fermi-like region compare to green area, a large positive $C_{\mathrm{UV}} \approx +5.2$
is required, in order to achieve a negative (enhanced fermionization regime) $\delta_u^{\text{eff}}$. If we have smaller $C_{\mathrm{UV}}$ then the green (Bose-like) area dominates.} 
                 \label{fig1} 
                 \end{figure}

\begin{itemize}
    \item $\delta_u - C_{\mathrm{UV}} > 0$: This requires $\mathcal{F}_{\mathrm{low}} > 0$, which occurs when low-energy modes are effectively enhanced. This corresponds to $g(\epsilon,\phi) > 1$ in the infrared, leading to bosonic behaviour.

    \item $\delta_u - C_{\mathrm{UV}} < 0$: This requires $\mathcal{F}_{\mathrm{low}} < 0$, which arises when low-energy modes are suppressed relative to higher-energy contributions. This corresponds to $g(\epsilon,\phi) < 1$ in the infrared, inducing an effective exclusion principle that leads to fermionization \ref{fig1}.

\end{itemize}

Importantly, fermionization does not occur simply because $g<1$, but because $g(\epsilon,\phi)$ creates a competition between low energy suppression and high energy enhancement. This competition is realized inside the thermal window $\phi \in [\pi/3,\,5\pi/3]$, where the kernel changes its relative weighting across energy scales. Outside this window, $g(\epsilon,\phi)>1$ for all energies, and the system reduces to a conventional bosonic regime without fermionization effects. We see the phase structure of the model in (\ref{fig1}). Notably, we see that lowering $T/t$ means thermal fluctuations are reduced. Inside the thermal window, the low-energy modes dominate the gap equation. These low-energy modes push $\delta_u$ negative, favoring $U>U_c$ (strong repulsion), which is the fermionization-like regime. In contrast, at higher temperatures, thermal excitations push $\delta_u$ positive, favoring Bose-like behaviour.

Interestingly, the condition $e^{-x}=\frac{1}{2}$ plays a role analogous to the appearance of $\ln 2$ in the $CP^{N-1}$ model at imaginary chemical potential \cite{Filothodoros, Filothodoros2}, but in our case, while in the continuous model this configuration appears as the free energy density, in our formulation it emerges through the structure of the thermal kernel.

\subsection{Extrema of $N_b(\phi)$ at $\phi=\pi/3,\,5\pi/3$}

We analyze the extrema and curvature of the boson number $N_b(\phi)$
for the repulsive Bose--Hubbard model at imaginary chemical potential.
Using the Bose distribution identity,
\begin{equation}
f_B(E - i\phi) - f_B(E + i\phi)
=
\frac{\sin\phi}{\cosh(\beta E) - \cos\phi},
\end{equation}
the boson number at $\Phi=0$ reads
\begin{equation}
N_b(\phi,T)
=
\int_{-2t}^{2t} d\epsilon \, \rho(\epsilon)\,
\frac{\sin\phi}{\cosh(\beta|\epsilon|) - \cos\phi}.
\label{eq:Nb}
\end{equation}
Assuming band symmetry,
\begin{equation}
\rho(\epsilon)=\rho(-\epsilon),
\end{equation}
the integrand is an even function of $\epsilon$.
If we define
\begin{equation}
I_B(\phi)
=
\int_{-2t}^{2t} d\epsilon \, \rho(\epsilon)\,
\frac{1}{\cosh(\beta|\epsilon|) - \cos\phi},
\end{equation}
and
\begin{equation}
J_B(\phi)
=
\int_{-2t}^{2t} d\epsilon \, \rho(\epsilon)\,
\frac{1}{\bigl(\cosh(\beta|\epsilon|) - \cos\phi\bigr)^2}
\end{equation}
then the first derivative is
\begin{equation}
\frac{dN_b}{d\phi}
=
\cos\phi \, I_B(\phi)
+
\sin^2\phi \, J_B(\phi).
\label{eq:dNb}
\end{equation}
Stationary points are determined by
\begin{equation}
\cos\phi \, I_B(\phi)
+
\sin^2\phi \, J_B(\phi)
= 0.
\label{eq:extremum_boson}
\end{equation}
At the special angles $\phi = \frac{\pi}{3}, \frac{5\pi}{3}$,
we have
$\cos\phi = \frac{1}{2},\sin^2\phi = \frac{3}{4}$
so Eq.~(\ref{eq:extremum_boson}) yields
\begin{equation}
\frac{1}{2} I_B + \frac{3}{4} J_B = 0
\quad \Longrightarrow \quad
J_B = -\frac{2}{3} I_B.
\label{eq:JB_relation}
\end{equation}
Differentiating Eq.~(\ref{eq:dNb}) gives
\begin{equation}
\frac{d^2 N_b}{d\phi^2}
=
-\sin\phi \, I_B
+ \cos\phi \, I_B'
+ 2\sin\phi\cos\phi \, J_B
+ \sin^2\phi \, J_B'.
\end{equation}
and using the identity
\begin{equation}
I_B'(\phi) = -\sin\phi \, J_B(\phi),
\end{equation}
we obtain
\begin{equation}
\frac{d^2 N_b}{d\phi^2}
=
-\sin\phi \, I_B
+ \sin\phi\cos\phi \, J_B
+ \sin^2\phi \, J_B'.
\label{eq:d2Nb_general}
\end{equation}
Using
$\sin\phi = \frac{\sqrt{3}}{2},
\cos\phi = \frac{1}{2}$
Eq.~(\ref{eq:d2Nb_general}) gives
\begin{equation}
\frac{d^2 N_b}{d\phi^2}\bigg|_{\phi=\pi/3}
=
-\frac{\sqrt{3}}{2} I_B
+ \frac{\sqrt{3}}{4} J_B
+ \frac{3}{4} J_B'.
\end{equation}
From Eq.~(\ref{eq:JB_relation}),
we obtain
\begin{equation}
\boxed{
\frac{d^2 N_b}{d\phi^2}\bigg|_{\phi=\pi/3}
=
-\frac{2\sqrt{3}}{3}\, I_B
+ \frac{3}{4}\, J_B'.
}
\label{eq:d2Nb_pi3}
\end{equation}
The derivative $J_B'(\phi)$ is given by
\begin{equation}
J_B'(\phi)
=
-2\sin\phi
\int_{-2t}^{2t} d\epsilon \, \rho(\epsilon)\,
\frac{1}{\bigl(\cosh(\beta|\epsilon|) - \cos\phi\bigr)^3}.
\end{equation}
At $\phi = \pi/3$, $\sin\phi > 0$, and the integrand is positive,
so
\begin{equation}
J_B'(\pi/3) < 0
\end{equation}
and since $I_B>0 \rightarrow \frac{d^2 N_b}{d\phi^2}\bigg|_{\phi=\pi/3}<0$.

At $\phi = 5\pi/3$ we have $
\sin\phi = -\frac{\sqrt{3}}{2},
\cos\phi = \frac{1}{2}$
leading to
\begin{equation}
\boxed{
\frac{d^2 N_b}{d\phi^2}\bigg|_{\phi=5\pi/3}
=
\frac{2\sqrt{3}}{3}\, I_B
+ \frac{3}{4}\, J_B'.
}
\label{eq:d2Nb_5pi3}
\end{equation}
In this case, since $\sin\phi < 0$, one finds
\begin{equation}
J_B'(5\pi/3) > 0
\end{equation}
and since $I_B>0 \rightarrow \frac{d^2 N_b}{d\phi^2}\bigg|_{\phi=5\pi/3}>0$.
So:

\begin{itemize}
\item At $\phi = \pi/3$: $\displaystyle \frac{dN_b}{d\phi}=0$ and
$\displaystyle \frac{d^2N_b}{d\phi^2}<0$, therefore $N_b$ has a
\emph{local maximum}. This corresponds to the maximal particle-hole
asymmetry with particle-like excitations dominating in fermionization regime.
\item At $\phi = 5\pi/3$: $\displaystyle \frac{dN_b}{d\phi}=0$ and
$\displaystyle \frac{d^2N_b}{d\phi^2}>0$, therefore $N_b$ has a
\emph{local minimum}. This corresponds to the maximal (negative)
particle-hole asymmetry with hole-like excitations dominating in fermionization regime.
\end{itemize}
This situation is corresponding to the fermionic case. At the edges of thermal window we have particle-like (maximum) at $2\pi/3$ and hole-like (minimum) at $4\pi/3$. On the other hand for bosons, at the edges of thermal window we have particle-like (maximum) at $\pi/3$ and hole-like (minimum) at $5\pi/3$.

\section{Notes on Attractive Fermi-Hubbard Model at Large-\(N\) Formulation}

In this section it is important to recall the physics of the main results from the attractive Fermi--Hubbard model \cite{Filothodoros3} with an imaginary chemical potential $\mu = i\theta$, which acts as a phase twist in imaginary time and modifies the Matsubara frequencies as $i\omega_n \rightarrow i\omega_n +i\theta$.

The inverse fermionic Green's function in Nambu space takes the form
\begin{equation}
\mathcal G^{-1}(i\omega_n,\mathbf k)=
\begin{pmatrix}
(i\omega_n +i\theta)-\epsilon_{\mathbf k} & \Delta \\
\Delta^* & (i\omega_n +i\theta)+ \epsilon_{\mathbf k}
\end{pmatrix},
\end{equation}
which preserves particle--hole symmetry.
The quasiparticle spectrum is determined by
\begin{equation}
E_{\mathbf k} = \sqrt{\epsilon_{\mathbf k}^2 + |\Delta|^2},
\end{equation}
and the determinant of the Green's function is
\begin{equation}
\det \mathcal G^{-1} =
(i\omega_n +i\theta)^2 - E_{\mathbf k}^2.
\end{equation}
The thermodynamic potential is given by
\begin{equation}
\Omega =
\frac{|\Delta|^2}{2U}
- \int \frac{dk}{2\pi} E_k
- \frac{1}{\beta} \int \frac{dk}{2\pi}
\ln\!\left[
1 + 2e^{-\beta E_k}\cos\phi + e^{-2\beta E_k}
\right],
\end{equation}
with $\phi = \beta\theta$.
The gap equation becomes
\begin{equation}
\frac{1}{U}
=
\int \frac{dk}{2\pi}
\frac{1}{E_k}
\left[
1 - f(E_k+i\theta) - f(E_k-i\theta)
\right],
\end{equation}
where $f(x)=1/(e^{\beta x}+1)$ is the Fermi--Dirac distribution.
The number equation reads
\begin{equation}
N_f =
\int \frac{dk}{2\pi}
\left[
f(E_k - i\theta) - f(E_k + i\theta)
\right],
\end{equation}
which measures the imbalance between particle-like and hole-like excitations induced by the imaginary chemical potential.
The thermodynamics is governed by the fermionic thermal kernel
\begin{equation}
g_F(x,\phi) = \frac{\sinh x}{\cosh x + \cos\phi}, \qquad x = \beta E_k.
\end{equation}
The kernel satisfies
\begin{equation}
g_F(x,\phi) - 1 =
- \frac{e^{-x} + \cos\phi}{\cosh x + \cos\phi},
\end{equation}
which determines the sign structure of contributions to the gap equation.
The special angles $\phi = 2\pi/3$ and $\phi = 4\pi/3$ (where $\cos\phi = -1/2$) define a universal thermal window in which the kernel changes sign at $x_* = \ln 2$.

Inside this thermal window, spectral weight is redistributed between low- and high-energy modes, leading to enhanced pairing correlations and driving the BCS--BEC crossover.
The crossover is controlled by the parameter
\begin{equation}
\delta_u = \frac{1}{U} - \frac{1}{U_c},
\end{equation}
where $\delta_u < 0$ corresponds to the BEC regime (strong attraction) and $\delta_u > 0$ to the BCS regime (weak attraction).
The imaginary chemical potential acts as a \textit{statistical regulator}, continuously tuning the system between different pairing regimes by modifying the entropy contribution to the free energy.
At the critical point $\delta_u = 0$, the system exhibits universal behaviour characterized by a thermal window which plays a central role in determining the phase structure.

\section{Mapping Between Attractive Fermi and Repulsive Bose Hubbard Models at Imaginary Chemical Potential}

\subsection{Formal Mapping via Matsubara Frequencies}

The partition function for fermions with imaginary chemical potential $i\theta$ is
\begin{equation}
\mathcal{Z}_F(\theta) = \prod_{\mathbf{k},n} \left[ (i\omega_n^F +i\theta)^2 - E_{\mathbf{k}}^2 \right],
\qquad \omega_n^F = \frac{(2n+1)\pi}{\beta}.
\end{equation}
For bosons, the Matsubara frequencies are $\omega_n^B = 2n\pi/\beta$.  Using the relation
$\omega_n^B = \omega_n^F + \pi/\beta$, we obtain
\begin{equation}
i\omega_n^B +i\theta = i\omega_n^F +i(\theta +\pi).
\end{equation}
Consequently,
\begin{equation}
\boxed{ \mathcal{Z}_B(\theta) = \mathcal{Z}_F(\theta +\pi) }.
\end{equation}
Thus, the repulsive Bose-Hubbard model at imaginary chemical potential $\theta$ is 
exactly dual to the attractive Fermi-Hubbard model at $\theta +\pi$.
Within the fermionization-like window, the free energy $F = U - TS$ is minimized by interaction avoidance since the phase twist increases effective density pressure.
Then bosons are rearranged into higher energy levels, mimicking Pauli exclusion to avoid
the repulsive interaction cost $U$.
Also entropic gain makes the fermionized distribution occupies a larger spectral range
than a standard Bose-Einstein condensate, lowering $F$.

\subsection{Mapping Between Fermionic and Bosonic Thermal Kernels}

The fermionic thermal kernel derived from the attractive Hubbard model is
\begin{equation}
g_F(x,\phi) = \frac{\sinh x}{\cosh x + \cos\phi}, \qquad x = \beta E_k, \quad \phi = \beta\theta.
\end{equation}

For the repulsive Bose-Hubbard model, the corresponding bosonic thermal kernel is
\begin{equation}
g_B(x,\phi) = \frac{\sinh x}{\cosh x - \cos\phi}.
\end{equation}

The two kernels are related by a simple shift of the imaginary chemical potential:
\begin{equation}
\boxed{ g_B(x,\phi) = g_F(x,\phi + \pi) }.
\end{equation}

Indeed, using $\cos(\phi + \pi) = -\cos\phi$, we obtain
\begin{equation}
g_F(x,\phi + \pi) = \frac{\sinh x}{\cosh x + \cos(\phi + \pi)} = \frac{\sinh x}{\cosh x - \cos\phi} = g_B(x,\phi).
\end{equation}
This mapping reflects the underlying duality between the attractive Fermi and repulsive Bose Hubbard models at imaginary chemical potential, with the shift $\phi \to \phi + \pi$ exchanging the roles of the two statistics.

\section{Universal Statistical Transmutation Framework at Imaginary Chemical Potential}

We consider a generic interacting lattice system at finite temperature
with an imaginary chemical potential $\mu =i\theta$.
The grand-canonical partition function can be written as
\begin{equation}
\boxed{
\mathcal{Z}(\phi)
=
\prod_{\mathbf{k},n}
\left[
(i\omega_n +i\theta)^2 - E_{\mathbf{k}}^2
\right]^{\sigma},}
\end{equation}
where $\sigma = +1$ for fermions and $\sigma = -1$ for bosons.
Taking the logarithm and performing the Matsubara sum yields the universal free energy density:
\begin{equation}
\boxed{
\Omega_\sigma(\phi)
=\frac{\gamma_{\sigma}^2}{2U}
- \int \frac{d^d k}{(2\pi)^d}
\left[
E_{\mathbf{k}}
+
\frac{\sigma}{\beta}\, \ln\left(
1 + \sigma\, 2 e^{-\beta E_{\mathbf{k}}} \cos\phi
+ e^{-2\beta E_{\mathbf{k}}}
\right)
\right].
}
\end{equation}
where $\gamma_{+1}\equiv \Delta$ and $\gamma_{-1}\equiv\Phi$ and this expression unifies fermionic and bosonic statistics through the parameter $\sigma$.
Differentiating the free energy with respect to $E$ yields the universal occupation kernel:
\begin{equation}
\boxed{
g_\sigma(x,\phi)
=
\frac{\sinh x}{\cosh x + \sigma \cos\phi},
\qquad x = \beta E.
}
\end{equation}
This kernel governs all thermodynamic observables.

\paragraph{Fermions ($\sigma=+1$):}
\begin{equation}
g_F(x,\phi)
=
\frac{\sinh x}{\cosh x + \cos\phi}
\end{equation}

\paragraph{Bosons ($\sigma=-1$):}
\begin{equation}
g_B(x,\phi)
=
\frac{\sinh x}{\cosh x - \cos\phi}
\end{equation}
and so the free energy densities obey
\begin{equation}
\boxed{
\Omega_B(\phi) = \Omega_F(\phi +\pi).
}
\end{equation}
This establishes a precise duality:
\begin{align*}
\boxed{
\textit{Bosonic thermal kernel at phase $\phi$  is equivalent to fermionic thermal kernel at phase $\phi+\pi$.}
}
\end{align*}
The kernel satisfies
\begin{equation}
g_\sigma(x,\phi) - 1
=
-\frac{e^{-x} + \sigma \cos\phi}{\cosh x + \sigma \cos\phi}.
\end{equation}
The sign change occurs at
\begin{equation}
e^{-x_*} = -\sigma \cos\phi.
\end{equation}
Thus:
\begin{itemize}
\item Fermions ($\sigma=+1$): critical angle $\phi = 2\pi/3$
\item Bosons ($\sigma=-1$): critical angle $\phi = \pi/3$
\end{itemize}
This reveals a universal $\mathbb{Z}_3$ structure:
\begin{equation}
\boxed{
\phi_c^{(B)} = \phi_c^{(F)} +\pi.
}
\end{equation}
The order parameter $\Delta$ (fermions) or $\Phi$ (bosons) satisfies:
\begin{equation}
\boxed{
\frac{1}{U} - \frac{1}{U_c}
=
\int d\epsilon\, \rho(\epsilon)
\left[
\frac{g_\sigma(\beta E,\phi)}{E}
- \frac{1}{\epsilon}
\right].
}
\end{equation}
This single equation describes the BCS--BEC crossover for fermions, the fermionization-like crossover for bosons and dimensional effects via $\rho(\epsilon)$.
Interestingly, after this map between the two theories, we may say that imaginary chemical potential which shifts Matsubara frequencies, modifies occupation statistics, and creates effective exclusion (fermionization). The free energy then becomes:
\begin{equation}
F = U - TS(\phi),
\end{equation}
where $\phi$ controls the entropy functional.

On the other hand, if one wants to find the critical thermal window at unitarity then using the Landau theory of phase transitions by expanding the thermal potential near a critical point, then the phase boundary is determined by:
\begin{equation}
a_2(T,\phi) = 0
\end{equation}
with
\begin{equation}
a_2 =
\delta_u-\int d\epsilon\, \rho(\epsilon)
\frac{g_\sigma(\beta\epsilon,\phi) - 1}{\epsilon}.
\end{equation}
arising from analytical calculation in (\ref{Appendix}).
It is shown in \ref{tab:duality} the duality picture where BCS--BEC crossover and bosonic fermionization arise from the same underlying mechanism.
\begin{table}[h]
\centering
\caption{Duality between attractive fermions and repulsive bosons at imaginary chemical potential.}
\label{tab:duality}
\small
\begin{tabular}{|p{4cm}|c|c|}
\hline
\textbf{Feature} & \textbf{Fermions (Attractive)} & \textbf{Bosons (Repulsive)} \\ \hline
Interaction & $-|U|$ (attractive) & $+U$ (repulsive) \\
Thermal window & $2\pi/3 < \phi < 4\pi/3$ & $\pi/3 < \phi < 5\pi/3$ \\
Inside behavior & BCS/BEC possible & Fermionization possible \\ \hline
\multicolumn{3}{|c|}{\textbf{Low temperature ($T/t< 1$)}} \\ \hline
Inside window & BEC-like (more pairing) & Fermi-like (more spreading) \\
Outside window & Normal fermions & Pure bosons \\ \hline
\multicolumn{3}{|c|}{\textbf{High temperature ($T/t>1$)}} \\ \hline
Inside window & Pairing fluctuations suppressed & Bose-like (less spreading) \\
Outside window & Normal fermions & Standard Bose-enhanced regime \\ \hline
\multicolumn{3}{|c|}{\textbf{Coupling strength ($\delta_u = 1/U - 1/U_c$)}} \\ \hline
$\delta_u < 0$ ($U > U_c$) & More pairing (BEC) & More spreading (Fermi-like) \\
$\delta_u > 0$ ($U < U_c$) & Less pairing (BCS) & Less spreading (Bose-like) \\
$\delta_u = 0$ & Unitarity (critical) & Fermionization point (critical) \\ \hline
\end{tabular}
\end{table}

\section{Discussion}

We have established an exact duality between the attractive Fermi-Hubbard model and the repulsive Bose-Hubbard model at finite temperature in the presence of an imaginary chemical potential $\mu = i\theta$. Using the relation $\omega_n^B = \omega_n^F + \pi/\beta$ between bosonic and fermionic Matsubara frequencies, we find $\mathcal{Z}_B(\theta) = \mathcal{Z}_F(\theta + \pi)$, which maps the repulsive Bose-Hubbard model onto the attractive Fermi-Hubbard model shifted by $\pi$. Consequently, the thermal kernels satisfy $g_B(x,\phi) = g_F(x,\phi + \pi)$ with $\phi = \beta\theta$, and the unified kernel $g_\sigma(x,\phi) = \sinh x/(\cosh x + \sigma \cos\phi)$ ($\sigma = +1$ for fermions, $\sigma = -1$ for bosons) governs both models. The condition $g_\sigma(x,\phi) = 1$ yields $x_* = \ln 2$ when $\cos\phi = -\sigma/2$, which interestingly defines the universal thermal window boundaries: $\phi = 2\pi/3,4\pi/3$ for fermions and $\phi = \pi/3,5\pi/3$ for bosons. Inside these windows, spectral weight is redistributed between lower- and higher-energy modes. 

For fermions inside the window, $\delta_u = 1/U - 1/U_c < 0$ (strong attraction) favors BEC-like pairing, while $\delta_u > 0$ (weak attraction) favors BCS-like behaviour. For bosons inside the fermionization window, $\delta_u < 0$ (strong repulsion) favors Fermi-like spreading, while $\delta_u > 0$ (weak repulsion) favors Bose-like behaviour. In both cases, lowering $T/t$ enhances the special behaviour inside the window, as the entropy term $-TS$ becomes less dominant and the interaction energy governs the free energy $F = U - TS$. The imaginary chemical potential thus acts as a temporal gauge field that continuously interpolates between Bose and Fermi statistics, with dimensionality entering only through the density of states $\rho(\epsilon)$.
 
We think that our results offer a new window into the physics of fermionization and Bose-Fermi mapping at finite temperature since this duality unifies BCS-BEC crossover in fermions and fermionization in bosons within a single framework. Notably, like the fermionic case, where the extrema of fermion number $N_f(\phi)$ are fixed by particle--hole symmetry, the extrema of boson number $N_b(\phi)$ are also universal. This is a result which makes the curvature exhibit a dynamic selection of particle- or hole-like behaviour at the window boundaries which they correspond to fixed maxima/minima.

Although boson–fermion mappings at imaginary chemical potential have been explored in field-theoretic contexts \cite{Filothodoros,Filothodoros2}, the interplay between imaginary chemical potential and fermionization in lattice bosonic and fermionic systems remains largely unexplored. Our findings imply that they could be extended in several new directions and present a remarkable progress in our understanding
of duality identities, like an overall Fermi-Boson Hubbard model coupled to a gauge field or it would be interesting to examine our theory with experimental results arising from the appearance of synthetic gauge fields or Floquet engineering \cite{ Scholl}. This could be an opened window to ultracold atoms like \cite{Bonkhoff2} where anyon statistical angle is the closest experimental analogue to our $\phi$ from imaginary chemical potential. Also a Bose-Fermi Hubbard map for general $d$ dimensions would be an interesting concept along with the fermionization-like windows as a generalization to the $1D$ results or examine if the edges of thermal windows are at different angles determined by the zeros or extrema of Clausen functions.

\section*{Appendix. The $a_2=0$ critical point for general statistics}
\label{Appendix}
If one wants to find the critical thermal window at unitarity then using the Landau theory of phase transitions by expanding the thermal potential near a critical point, then:
\begin{equation}
\Omega(\gamma) = \Omega(0) + a_2 \gamma^2 + a_4 \gamma^4 + a_6 \gamma^6 + \cdots,
\end{equation}
where $a_2 = 0$ defines the critical point where the curvature of the thermodynamic potential changes sign and $\gamma$ is the fermionic/bosonic parameter ($\Delta$ or $\Phi$).

We start from the unified thermodynamic potential for a generic interacting lattice system with imaginary chemical potential,
\begin{equation}
\Omega(\gamma) = \frac{\gamma^2}{2U}
-\Big[\int \frac{dk}{2\pi} E_k
+\frac{\sigma}{\beta} \int \frac{dk}{2\pi} \ln\!\left[1 + \sigma\, 2e^{-\beta E_k}\cos\phi + e^{-2\beta E_k}\right]\Big],
\label{eq:Omega_start_general}
\end{equation}
where $E_k = \sqrt{\epsilon_k^2 + \Delta^2}$, $\phi = \beta\theta$, and $\sigma = +1$ for fermions, $\sigma = -1$ for bosons.

The first derivative with respect to $\gamma$ is
\begin{equation}
\frac{\partial \Omega}{\partial \gamma}
= \frac{\gamma}{U}
- \int \frac{dk}{2\pi} \frac{\gamma}{E_k}
- \frac{\sigma}{\beta} \int \frac{dk}{2\pi} \frac{1}{X_\sigma} \frac{\partial X_\sigma}{\partial \gamma},
\end{equation}
with $X_\sigma = 1 + \sigma\, 2e^{-\beta E_k}\cos\phi + e^{-2\beta E_k}$.

We have
\begin{equation}
\frac{\partial X_\sigma}{\partial \gamma}
= -2\beta \frac{\gamma}{E_k} e^{-\beta E_k}\cos\phi
- 2\beta \frac{\gamma}{E_k} e^{-2\beta E_k}
= -2\beta \frac{\gamma}{E_k} e^{-\beta E_k}\bigl(\cos\phi + e^{-\beta E_k}\bigr).
\end{equation}
Note that the derivative is independent of $\sigma$ because $\sigma$ multiplies only the $\cos\phi$ term in $X_\sigma$, which vanishes upon differentiation with respect to $\gamma$.

Substituting,
\begin{equation}
\frac{\partial \Omega}{\partial \gamma}
= \frac{\gamma}{U} - \int \frac{dk}{2\pi} \frac{\gamma}{E_k}
+\sigma \int \frac{dk}{2\pi} \frac{\gamma}{E_k}
\frac{e^{-\beta E_k}\bigl(\cos\phi + e^{-\beta E_k}\bigr)}{X_\sigma}.
\end{equation}
Using the identity
\begin{equation}
X_\sigma = 1 + \sigma\, 2e^{-\beta E_k}\cos\phi + e^{-2\beta E_k}
= 2e^{-\beta E_k}\bigl(\cosh(\beta E_k) + \sigma \cos\phi\bigr),
\end{equation}
we obtain
\begin{equation}
\frac{e^{-\beta E_k}\bigl(\cos\phi + e^{-\beta E_k}\bigr)}{X_\sigma}
= \frac{\cos\phi + e^{-\beta E_k}}{2\bigl(\cosh(\beta E_k) + \sigma \cos\phi\bigr)}.
\end{equation}
Thus,
\begin{equation}
\frac{\partial \Omega}{\partial \gamma}
= \frac{\gamma}{U} - \int \frac{dk}{2\pi} \frac{\gamma}{E_k}
+ \sigma \int \frac{dk}{2\pi} \frac{\gamma}{E_k}
\frac{\cos\phi + e^{-\beta E_k}}{\cosh(\beta E_k) + \sigma \cos\phi}.
\label{eq:first_derivative_general}
\end{equation}
The second derivative is
\begin{equation}
\frac{\partial^2 \Omega}{\partial \gamma^2}
= \frac{1}{U} - \int \frac{dk}{2\pi} \frac{1}{E_k}
+ \sigma \int \frac{dk}{2\pi} \frac{1}{E_k}
\frac{\cos\phi + e^{-\beta E_k}}{\cosh(\beta E_k) + \sigma \cos\phi}
+ \gamma \frac{\partial}{\partial \gamma} \bigl[ \cdots \bigr],
\end{equation}
where the term proportional to $\gamma$ vanishes as $\gamma \to 0$. Hence, the Landau coefficient $a_2(T,\phi) = \partial^2 \Omega / \partial \gamma^2|_{\gamma\rightarrow 0}$ is
\begin{equation}
a_2(T,\phi) = \frac{1}{U} - \int \frac{dk}{2\pi} \frac{1}{E_k}
+ \sigma \int \frac{dk}{2\pi} \frac{1}{E_k}
\frac{\cos\phi + e^{-x}}{\cosh x + \sigma \cos\phi},
\label{eq:a2_final_general}
\end{equation}
where $x = \beta E_k$.

Since 
\begin{equation}
\frac{\cos\phi + e^{-x}}{\cosh x + \sigma\cos\phi}=1-\frac{\sinh x}{\cosh x + \sigma\cos\phi}
\end{equation}
and after inserting $-1/U_c$ and $+1/U_c$ we take finally
\begin{equation}
a_2(T,\phi) = \delta_u-\int_{\mathrm{BZ}} \frac{dk}{2\pi} \left[ 
\frac{\sinh(\beta E_k)}{E_k\big[\cosh(\beta E_k)+\sigma\cos\phi\big]}
- \frac{1}{|\epsilon_k|}
\right]
\end{equation}
where at $a_2=0$ we find the critical point where the thermal potential changes sign which gives
\begin{equation}
\boxed{
\delta_u = \int d\epsilon \, \rho(\epsilon) \,
\frac{g_\sigma(\beta\epsilon,\phi) - 1}{\epsilon}
}.
\label{eq:critical_condition}
\end{equation}

This unified formulation shows that the entire phase structure for both fermions and bosons is encoded in the single kernel $g_\sigma(x,\phi)$, with the statistics controlled by the sign $\sigma$
where at $a_2=0$ we find the critical point where the thermal potential changes sign.

\ack{I would like to thank Anastasios Petkou for helpful discussion.}

\data{All data that support the findings of this study are included within the article.}

\section*{Conflict of Interest Statement}
The author declares no conflict of interest.

\section*{Funding Statement}
This research received no external funding.

\end{document}